\documentclass{SCIS2016}
\usepackage{cite}
\usepackage{geometry}

\begin{document}
%%%%%%%%%%%%%%%%%%%%%%%%%%%%%%%%%%%%%%%%%%%%%%%%%%%%%%%
%%% Authors do not modify the information below
%%% 作者不需要修改此处信息
%%% 有专题名称时, 将第一行的{}注释掉, 使用第二行

\ArticleType{SURVEY PAPER}{}
%{\put(0,-26){}}
\Year{2016}
\Month{January}
\Vol{59}
\No{1}
\DOI{xxxxxxxxxxxxxx}
\ArtNo{xxxxxx}
\ReceiveDate{}
\AcceptDate{}
\OnlineDate{}
%%%%%%%%%%%%%%%%%%%%%%%%%%%%%%%%%%%%%%%%%%%%%%%%%%%%%%%

%%% title: 标题
%%%%%%%%%%%%%%%%%%%%%%%%%%%%%%%%%%%%%%%%%%%%%%%%%%%%%%%
\title{Channel Measurements and Models for High-Speed Train Wireless Communication Systems in Tunnel Scenarios: A Survey}{Channel measurements and models for high-speed train wireless communication systems in tunnel scenarios: a survey}
%%% Corresponding author: 通信作者
%%%   \author[number]{Full name}{{email@xxx.com}}
%%% General author: 一般作者
%%%   \author[number]{Full name}{}
\author[1]{Yu Liu}{}
\author[2]{Ammar Ghazal}{}
\author[3]{Cheng-Xiang Wang}{cheng-xiang.wang@hw.ac.uk}
\author[4]{Xiaohu Ge}{}
\author[5]{Yang Yang}{}
\author[1]{Yapei Zhang}{}
%%% Author information for page head. 页眉中的作者信息
%%% 若此处指定以此处为准, 否则直接调用author信息
\AuthorMark{Liu Y}
%%% Authors for citation. 首页引用中的作者信息
%%% 若此处指定以此处为准, 否则直接调用author信息
\AuthorCitation{Liu Y, Ghazal A, Wang C X, {\em et al}}
%%% Address. 地址
%%%   \address[number]{Address, City {\rm Postcode}, Country}
\address[1]{Shandong Provincial Key Lab of Wireless Communication Technologies, Shandong University, Jinan, {\rm 250100}, China}
\address[2]{Centre for Electronic and Communications Engineering, School of Engineering and Sustainable Development,\\ 
De~Montfort University, Leicester, {\rm LE1 9BH}, U.K.}
\address[3]{Institute of Sensors, Signals and Systems, School of Engineering \& Physical Sciences, Heriot-Watt University, \\ Edinburgh, {\rm EH14 4AS}, U.K.}
\address[4]{Department of Electronics and Information Engineering, Huazhong University of Science and Technology, \\
Wuhan, {\rm 430074}, China}
\address[5]{Key Laboratory of Wireless Sensor Network \& Communication, Shanghai Institute of Microsystem and \\
Information Technology (SIMIT), Chinese Academy of Sciences (CAS), Shanghai, {\rm 200050}, China}

%%% Abstract. 摘要
\abstract{The rapid developments of high-speed trains (HSTs) introduce new challenges to HST wireless communication systems. Realistic HST channel models play a critical role in designing and evaluating HST communication systems. Due to the length limitation, bounding of tunnel itself, and waveguide effect, channel characteristics in tunnel scenarios are very different from those in other HST scenarios. Therefore, accurate tunnel channel models considering both large-scale and small-scale fading characteristics are essential for HST communication systems. Moreover, certain characteristics of tunnel channels have not been investigated sufficiently. This article provides a comprehensive review of the measurement campaigns in tunnels and presents some tunnel channel models using various modeling methods. Finally, future directions in HST tunnel channel measurements and modeling are discussed.}

%%% Keywords. 关键词.
\maketitle
%%%%%%%%%%%%%%%%%%%%%%%%%%%%%%%%%%%%%%%%%%%%%%%%%%%%%%%
%%% The main text. 正文部分
%%%%%%%%%%%%%%%%%%%%%%%%%%%%%%%%%%%%%%%%%%%%%%%%%%%%%%%
\section{Introduction}\label{Sec1}
High-speed trains (HSTs) have experienced a rapid development recently~\cite{IMT_5G}, and wireless technologies for HST communications will be considered as an important issue in the fifth generation (5G) wireless communication networks \cite{Wan14_5G}.
With the number of HST users increasing, numerous communication data need to be transmitted to train passengers through wireless channels. Therefore, high-capacity and reliable HST communication networks regardless of users' locations or speeds, are required. In order to meet these requirements, HST wireless communication systems need to mitigate various challenges resulting from the high speed of the train, such as frequent handovers, large Doppler spreads, and fast travel through different HST scenarios \cite{Ai_Scenarios12}. Currently, the most widely used HST communication system is the Global System for Mobile Communication Railway (GSM-R) \cite{Gha15_HST}. It can be used for the train communication and control. However, it cannot satisfy the increasing communication requirements of high data rates. In recent years, the Long-Term Evolution-Railway (LTE-R) system, which is based on the LTE-Advanced (LTE-A) system, has been recommended to replace the GSM-R. Both of them adopted the conventional network architecture in which HST users inside the train communicate with outdoor base stations (BSs) directly. However, such an architecture results in high penetration losses when the signals travel into the carriages and leads to a spotty coverage, which will result in the handover failure or the high drop calls rate \cite{Gha15_HST}. To overcome the above problems, the promising mobile relay station (MRS) solution has been proposed for future HST communication systems to solve the spotty coverage problem and mitigate high penetration losses of wireless signals traveling into the train carriages \cite{Fok10_Sur}. By considering the MRS technology, the propagation channel can be divided into two parts: outdoor channel between the BS and MRS, and indoor one between the MRS and receiver inside carriages. MRS technology can be applied to reduce the frequency handover by performing a group handover instead of individual handover with each passenger. This technology has been adopted by the International Mobile Telecommunications-Advanced (IMT-A) and WINNER~ II systems \cite{Wang16_HST_Sur}.

There are several scenarios that HST may encounter in reality, such as open space, hilly terrain, cutting, viaduct, tunnel, and station scenarios \cite{Ai_Scenarios12}. As a classical HST scenario, tunnel environment takes up a great proportion among the underground rail transportation, and it has attracted a lot of research interest \cite{Hro14_Sur}. Considering the unique propagation environment of tunnels, such as long limited space, poor smoothness of the interior walls, and the bounding of tunnel walls, propagation characteristics of signals in the tunnel scenario are quite different from those of other HST scenarios. Moreover, the length of tunnels can range from several hundred meters to several kilometers, with different shapes such as circular and rectangular sections. Since the length, size, and shape of the tunnels and the encountered waveguide phenomena have a significant effect on the propagation channel, channel characterization and modeling for tunnel scenarios are still a quite challenging topic \cite{Wang16_HST_Sur}.
Conventional channel modeling methods are suitable for most HST scenarios, but are not directly applicable to tunnel scenarios.
Radio signals inside tunnels will suffer more reflections, diffractions, and scattering, and will also experience more severe fast fading due to the high mobility of trains.
There are mainly two promising solutions to achieve the wireless coverage inside tunnels, i.e., leaky feeders \cite{Gua12_Leakey} and distributed antenna system (DAS) \cite {Bri07_DAS}, \cite{Gua13_DAS}. Leaky feeder solution was seen as a feasible technology to provide communication service inside tunnels. In this case, tunnel channel models are not needed for system design. However, considering the fact that HSTs may require long tunnels, leaky feeder technologies will be more expensive especially at high operating frequencies. Moreover, the leaky feeder will be unavailable once unexpected cuts appear \cite{Hro14_Sur}.
As a consequence, DAS is more viable than the leaky feeder \cite{Gua13_DAS} and it can provide better coverage, higher capacity, and easier maintenance after being installed. All these facts push the application of DAS inside tunnels and motivate the work on the HST tunnel channel modeling \cite{Gua12_Leakey}.

To better develop future HST tunnel communication systems, a comprehensive understanding of channel characteristics and accurate tunnel channel models are essential.
Some measurement campaigns inside tunnels have been conducted to investigate the underlying physical phenomenon of signal propagation environments. 
Most of the measurement works have focused on the large-scale fading characteristics, such as path loss (PL) and shadowing fading (SF), which are crucial in network deployment and optimization. The small-scale fading channel models also play an important role in system design and schemes test. Therefore, accurate tunnel channel models considering both large-scale and small-scale fading characteristics are
essential. Based on different modeling approaches, the presented tunnel channel models can be divided into deterministic and stochastic channel models. The deterministic channel models are mainly based the geometrical optical (GO) based theory {\cite{Mah74_GO, Por02_GO, Zha16_GO, Wan98_GO}}, the waveguide method {\cite{For14_WG, Dud05_WG, Did01_WG, Zha97_WG, Ems75_mine}}, and numerical methods for Maxwell equations {\cite{Ran12_Maxwell, Wan00_Maxwell, Taf05_FDTM}}. The stochastic models can be classified into geometry-based stochastic models (GBSMs) and non-geometrical stochastic models (NGSMs) \cite{Wan14_FSMM}.
The aim of the paper is to survey the recent advances in channel measurements and modeling  for HST tunnel scenarios and future directions.

The remainder of this paper is organized as follows. HST channel measurements in tunnel scenarios are presented in Section~\ref{Sec2}. In Section~\ref{Sec3}, some HST tunnel channel models are described. Future research directions in HST tunnel channel measurements and models are discussed in Section~\ref{Sec4}. Finally, conclusions are drawn in Section~\ref{Sec5}.

\section{HST Channel Measurements in Tunnel Scenarios}
\label{Sec2}
Due to the high speed of trains and tunnel space limitation, measurement campaigns are difficult to be carried out accurately inside HST tunnels. Although some HST channel measurements in tunnels have been conducted in recent years, this is still a very challenging task. Here, we will briefly review and classify some tunnel channel measurements \cite{Gua12_Leakey}, \cite{Bri07_DAS}, \cite{For14_WG}, \cite{Aik98_Mea, Gua15_Cur, Zha03_PL, Kim02_PL, He11_mea, Par08_LS, Lie03_MIMO, Cai16_Mea, Jia15_Mea, Zha15_Mea, Li15_Mea, Zha00_PL, Par08_RP, Kim03_PL, Bas14_Mea} according to the carrier frequency, tunnel parameters, antenna configuration, and channel statistics, as illustrated in Table~\ref{table_channel_measurement}.
Moreover, some important analysis of tunnel channel measurements are summarized as follows.
\begin{table}
	\center
	    \caption{Important tunnel channel measurements.}
    \begin{tabular}{ | c | c | c | c | c | c |}
    \hline
\textbf{Ref.} & \textbf{Freq.} & \textbf{Scenario} &
\textbf{$\begin {matrix} \textrm{Tunnel} \\
\textrm{ parameters} \end {matrix} $ }
&\textbf{$\begin {matrix} \textrm{Antenna} \\
\textrm{ Config.} \end {matrix} $ }
&\textbf{Channel Statistics}\\
\hline
%\textbf{Antenna Config}\\
%\textbf{Characteristics of Tunnel}
\cite{Gua12_Leakey} & 2.4 GHz
& $\begin {matrix} \textrm{Arched}\\\textrm{subway} \\
\textrm{tunnel} \end {matrix} $
& $ \begin {matrix}
\textrm{wide tunnel:}\\
\textrm{9.8 m*6.2 m,}\\
\textrm{narrow tunnel:}\\
\textrm{4.8 m*5.3 m,}
\end {matrix} $
& SISO
&  $ \begin {matrix}
\textrm{SF, PL,}\\
\textrm{fast fading,}\\
\textrm{LCF,}\\
\textrm{AFD}
\end {matrix} $\\
\hline
\cite{Bri07_DAS}& 900 MHz & $\begin {matrix} \textrm{Arched} \\\textrm{railway} \\ \textrm{tunnel} \end {matrix} $
& $ \begin {matrix}
\textrm{height: 5.4 m,}\\
\textrm{width: 10.7 m,}\\
\textrm{length: 4000 m}
\end {matrix} $
& SISO
& PL\\
\hline
\cite{Par08_LS} & 2.8-5 GHz
& $\begin {matrix} \textrm{Semicircular} \\
\textrm{railway} \\
\textrm{tunnel} \end {matrix} $
& $ \begin {matrix}
\textrm{diameter: 8.6 m,}\\
\textrm{height(max): 6.1 m,}\\
\textrm{length: 3336 m}
\end {matrix} $
& MIMO
&  $ \begin {matrix}
\textrm{PL,}\\
\textrm{PDF,}\\
\textrm{CDF}
\end {matrix} $\\
\hline
\cite{Lie03_MIMO} & 900 MHz
& $\begin {matrix} \textrm{Arched} \\
\textrm{subway} \\
\textrm{tunnel} \end {matrix} $
& $ \begin {matrix}
\textrm{two-track tunnel:}\\
\textrm{width: 8 m,}\\
\textrm{length: 200 m,}\\
\textrm{single-track tunnel:}\\
\textrm{width: 5 m,}\\
\textrm{length: 100 m}
\end {matrix} $
& MIMO
&  $ \begin {matrix}
\textrm{CIR,}\\
\textrm{Correlation coefficient}
\end {matrix}$\\
\hline
\cite{Cai16_Mea} & 2.1376 GHz
& $ \begin {matrix}
\textrm{Subway}\\
\textrm{tunnel}
\end {matrix}$
& length: 34 km
& MIMO
&  $\begin {matrix} \textrm{PDP, PL,} \\
\textrm{K factor,} \\
\textrm{delay spread} \end {matrix}$\\
\hline
\cite{Li15_Mea} &
$ \begin {matrix}
\textrm{2.4 GHz,}\\
\textrm{5 GHz}
\end {matrix}$
& $ \begin {matrix}
\textrm{Horse-shoe}\\
\textrm{shaped}\\
\textrm{subway}\\
\textrm{tunnel}
\end {matrix}$
& $ \begin {matrix}
\textrm{straight: 240 m,}\\
\textrm{curve: 140 m}
\end {matrix} $
& SISO
&  $\begin {matrix} \textrm{PL,} \\
\textrm{rms delay spread,} \\
\textrm{channel stationarity,} \\
\textrm{channel capacity} \end {matrix}$\\
\hline
\cite{Zha00_PL}& $\begin {matrix} \textrm{465 MHz,} \\
\textrm{820 MHz} \end {matrix}$
& $\begin {matrix} 
\textrm{Arched} \\
\textrm{underground} \\
\textrm{railway} \end {matrix} $
& $ \begin {matrix}
\textrm{floor width: 5.8 m,}\\
\textrm{height: 4 m,}\\
\textrm{length: 980 m}
\end {matrix} $
& SISO
& PL\\
\hline
\cite{Par08_RP}& $\begin {matrix} \textrm{450 MHz-} \\
\textrm{5 GHz} \end {matrix}$
& $\begin {matrix} 
\textrm{Arched} \\
\textrm{railway} \\
\textrm{tunnel} \end {matrix} $
& length: 3000 m
& SISO
& PL\\
\hline
\cite{Kim03_PL}& $\begin {matrix} \textrm{884 MHz-} \\
\textrm{2.45 GHz} \end {matrix}$
& $\begin {matrix} 
\textrm{Rectangular} \\
\textrm{railway} \\
\textrm{tunnel} \end {matrix} $
& $ \begin {matrix}
\textrm{width: 14.7 m,}\\
\textrm{height: 6.15 m,}\\
\textrm{length: 360 m}
\end {matrix} $
& SISO
& PL\\
\hline
\cite{Bas14_Mea}& 2.49-4 GHz
& 
$\begin {matrix} 
\textrm{Rectangular} \\
\textrm{tunnel} \end {matrix} $
&
$ \begin {matrix}
\textrm{wide tunnel:}\\
\textrm{2.4 m*3.1 m,}\\
\textrm{narrow tunnel:}\\
\textrm{2.4 m*5.2 m,}
\end {matrix} $
& MIMO
&  $\begin {matrix} \textrm{PL,} \\
\textrm{delay spread} \end {matrix}$\\
\hline
\multicolumn{6}{| l |}{$\begin {matrix} \textrm{PDF: probability density function;
CDF: cumulative density function;
LCR: level crossing rate;}\\
\textrm{
AFD: average fade duration;
CIR: channel impulse responses;
PDP: power delay profile}
\end {matrix} $ } \\
\hline
    \end{tabular}
	\label{table_channel_measurement}
\end{table}

\subsection{Measurement setup}
For HST tunnel channel measurements, most works have focused on single-input single-output (SISO) antenna configuration, e.g., in \cite{Bri07_DAS}, \cite{Gua12_Leakey}, and \cite{He11_mea}. To meet the increasing requirements of future high speed data transmission, the multiple-input multiple-output (MIMO) antenna configuration \cite{Par08_LS}, \cite{Lie03_MIMO} systems are indispensable. The benefits of increasing channel capacity by using the MIMO technology in tunnel scenario were examined and demonstrated in \cite{Lie03_MIMO}.
Therefore, more channel measurements using MIMO system in HST tunnel scenarios are necessary and significant. 

Most of the existing measurement campaigns have been conducted based on the GSM-R system. A typical measurement of the tunnels on the new HSTs in Spain was introduced in \cite{Bri07_DAS}. It is noteworthy to mention that the frequency bands of the GSM-R system reported are 876--880 MHz for the uplink and 921--925 MHz for the downlink, since the measurement was conducted in Europe. In China and India, different frequency bands of the GSM-R system, i.e., 885--889 MHz for the uplink and 930--934 MHz for the downlink, are in use. In \cite{Bri07_DAS}, the viability of using DAS was demonstrated. Two GSM-R BSs at the entrance and exit of the tunnel were used. Between these two BSs, there were three repeaters connected with each other using radio over fiber (RoF) technology. The measurement has taken into account the effects of tunnel propagation, including curves, trains passing from outside to inside, and the case of two trains inside the tunnel. In this paper, modal approach is used to calculate the signals propagation in straight tunnel, and ray tracing method is applied to calculate the extra attenuations in curves case and in the entrance and exit of the tunnel. 
When a train passes from inside to outside the tunnel, the signal wave can experience strong fading due to the change of wave impedance and diffraction effect. 
Moreover, two trains are used in the 
measurements. One train stopped at different positions, i.e., close to one transmitter and the center between two transmitters, while the other one passes by. The signal shadowing caused by blocking effect of two trains traveling inside tunnel is investigated.   
Furthermore, all the above cases were measured by using the iso-frequency and multi-frequency distributed transmitters solutions. For the iso-frequency configuration, all the transmitters inside tunnel had same frequency. The multi-frequency trial uses the different frequencies. Compared with the multi-frequency transmitter, the iso-frequency transmitter was shown to have improvements of signal-to-noise ratio (SNR) and received power \cite{Bri07_DAS}.
Considering that the GSM-R is mainly used for train control, rather than providing communications for passengers inside trains, and it cannot meet the high data rate requirements of future HST communications.

In \cite{Cai16_Mea}, the Universal Mobile Terrestrial System (UMTS) and LTE systems have been recommended to replace the GSM-R. In these systems, the wideband signals can be used to analyze the precise channel characteristics with enhanced time resolution. Using the extracted parameters from CIRs, such as the cluster delay, Doppler frequency spreads, $K$-factor and correlation among these parameters, intra-cluster characteristics are investigated. Moreover, a path loss model for the tunnel scenarios is obtained. In \cite{Jia15_Mea}, actual channel measurement based on the LTE system has been carried out at 1.89~GHz in a mountain tunnel. Some main propagation characteristics are investigated. In \cite{Zha15_Mea}, the measurement campaigns are conducted at carrier frequencies of 1~GHz and 2.45~GHz, according to the measurement configuration of the fourth generation (4G) systems in railway environments. It provides detailed tunnel channel information, and can be used to develop a broadband channel model for tunnel communication systems.

In summary, HST tunnel wideband channel measurement campaigns with MIMO system, as well as larger carrier frequency and bandwidth than GSM-R are needed for future developments of HST tunnel communication system.
\subsection{Large-scale vs. small-scale fading}
%For the better design of the future communication system in HST tunnel scenario, a comprehensive understanding of both large-scale and small-scale channel characteristics is vital. 
Most of the existing measurement campaigns for tunnels mainly focused on large-scale fading parameters. In \cite{Cai16_Mea},\cite{Zha00_PL, Par08_RP, Kim03_PL, Bas14_Mea, Sav15_Mea}, the PL has been investigated, and in \cite{Li16_PL}, the channel characteristics of different antenna setups are compared, such as PL exponent and SF. 
In \cite{He_TR}, the relation between Fresnel zone and path loss component $n$ was analyzed based on the two-ray model. It was demonstrated that the 1st Fresnel zone is the underlying factor that will influence the $n$. Then, in \cite{Hro10_4S}, the four-slope path loss channel model was proposed, by taking the free space propagation region and the extreme far region into additional consideration. All these results can be used to guide measurement campaigns and physical layer design for the communication systems in tunnel.

The small-scale fading channel models also play an important role in the analysis and design of wireless communication systems, such as error control coding, interleaving, and equalization algorithms \cite {Hro14_Sur}.
In~\cite{Par08_LS}, large-scale and small-scale fading characteristics deduced from measurement data in semicircular tunnel are presented. In the case of small-scale fading, a Rice distribution can fit measurement well, and a uniform distribution can match the phase of the electric field. Moreover, the $K$-factor is also analyzed. In \cite{Gua12_Leakey}, a measurement was carried out in a subway tunnel and some propagation characteristics, such as LCR and AFD, have been computed and discussed. 
In future tunnel channel measurements, more channel statistics related to small-scale fading in HST tunnel scenarios are needed.
\subsection{Far region vs. near region inside tunnel}
When radio waves propagate inside a tunnel, the tunnel channel can roughly be divided into two regions based on the so-called breakpoint, namely, the near region and far region \cite{Zha97_WG}, \cite{Gua13_BP}.
Different propagation regions need different channel models to describe. The statistical properties of tunnel channels, including PL, SF, and small-scale fading characteristics, are greatly different before and after the breakpoint. Therefore, accurate determination of the breakpoint is very important for future tunnel channel measurements and models.
Based on a measurement campaign conducted in a subway environment at 2.4 GHz in \cite{Gua12_Leakey}, some signal propagation characteristics on the breakpoint were discussed, and the propagation regions were analyzed from the following three perspectives.

\begin{itemize}
\item [1)]According to the modal theory, radio propagation can be decomposed into different waveguide modes. For example, a rectangular tunnel can be considered as an oversized waveguide. According to the operating frequency and cutoff frequency, the modes propagating inside a tunnel can be estimated. In \cite{Mar51_Cutoff}, the cutoff frequency for a rectangular tunnel can be expressed as
\begin{equation}
{f_T} = \frac{1}{{2\sqrt {{\mu _0}{\varepsilon _0}} }}\sqrt {{{(\frac{m}{a})}^2} + {{(\frac{n}{b})}^2}}
\end{equation}
where $a$ and $b$ represent the tunnel width and height, $m$ and $n$ denote the propagation modes in the horizontal and vertical directions, $\mu_0$ and $\varepsilon_0$ are permeability and permittivity in free space.
When the operating frequencies are higher than the cutoff frequency, the corresponding signals can propagate inside tunnel. In the near region, the field usually consists of many modes. However, with the increase of the distance between the transmitter and receiver, the higher-order modes of a signal experience stronger attenuation, and most of the higher-order modes are lost before the breakpoint. In the far region, i.e., after the breakpoint, the lowest-order mode is dominant.

\item[2)] According to the ray-tracing theory, the direct wave is dominant in the near region, while the reflected waves are dominant in the far region \cite{Kwo04_RT}.

\item[3)] From the statistics point of view, the received signal can be presented as the sum of the line-of-sight (LoS) component and diffuse components reflected by tunnel walls, ceiling, and the ground \cite{Gua12_Leakey}. In the near region, the received signal may contain a strong LoS component and therefore, the Ricean $K$-factor can be relatively large. In the far region, the Ricean $K$-factor can be small, and even the Rayleigh distribution may be considered when there is no LoS component.
\end{itemize}

\subsection{Typical propagation zones}
In general, the LoS component can be strong in conventional  HST communication scenarios. In tunnels, due to its leakproofness, the reflected rays can remain greatly and become more dominant in the received signal. Considering the long delay characteristics caused by the reflections in tunnels, tunnel channel measurements in different propagation zones should be conducted, and cluster delays should be deeply analyzed. There are roughly three propagation zones in tunnels, i.e., LoS, non-LoS (NLoS), and far-LoS (FLoS) zones. When the receiver is very close to the transmitter, the LoS propagation happens. When the train moves away from the transmitter, the LoS may disappear and the NLoS zone appears. When the train is far away from the transmitter, the FLoS zone can appear if there is no clear LoS component.
In \cite{He11_mea}, three-dimensional (3D) FCFs were obtained based on measurements conducted in the aforementioned three zones and PDPs were obtained. The PDP in the LoS zone matches the exponential distribution very well. With the increase of travel distance in the tunnel, the LoS component disappears and numerous reflections remain due to the obturation of tunnel. Those reflections will lead to delay clusters, which fit the generalized extreme value distribution very well in the NLoS zone and Johnson SB distribution in the FLoS zone. This phenomenon is quite unique for tunnel scenarios and should be taken into account when designing HST communication systems \cite{Kwo04_RT}.
In addition, considering the train passing by the transmitter, near-shadowing zone (NSZ) is observed in \cite{Gua12_Leakey} before the LoS zone. In this short zone, the LoS between the transmitter and receiver is blocked, and the multipath  propagation is dominate.
\subsection{Parameters influencing radio propagation inside tunnel}
The radio waves inside tunnel suffer more reflections, diffractions, and scattering. The parameters, such as tunnel size, tunnel shape, internal electromagnetic (EM) properties of tunnel walls,  surface roughness, and antenna radiation and position, will affect the radio signal propagation inside tunnel \cite{Kim03_PL}.

Tunnel cross section has a distinctive influence on the propagation attenuation, especially with the increasing of the signal frequency \cite{Mol09_TS}.
There are different shapes of cross sections in real tunnels, such as circular, semicircular, rectangular, arched and oval ones. Some typical tunnel shapes are shown in Fig. 1. Due to the consideration of tunnel structure and construction, circular and arched tunnels are more often than rectangular ones. In \cite{Did01_TS}, a measurement campaign carried out in a subway tunnel has been introduced to characterize EM propagation in underground rail tunnel. It demonstrates that tunnel geometry, i.e., the shape of tunnel cross section and curves, have an important impact on the signal propagation, rather than the electromagnetic properties of materials. In \cite{Che10_TS} and \cite{Wan10_TS}, the influence of rectangular cross section in tunnel has been investigated, and in \cite{Cha10_TS} the attenuations in different tunnel shapes are analyzed.

\begin{figure}[!t]
\centering
\includegraphics{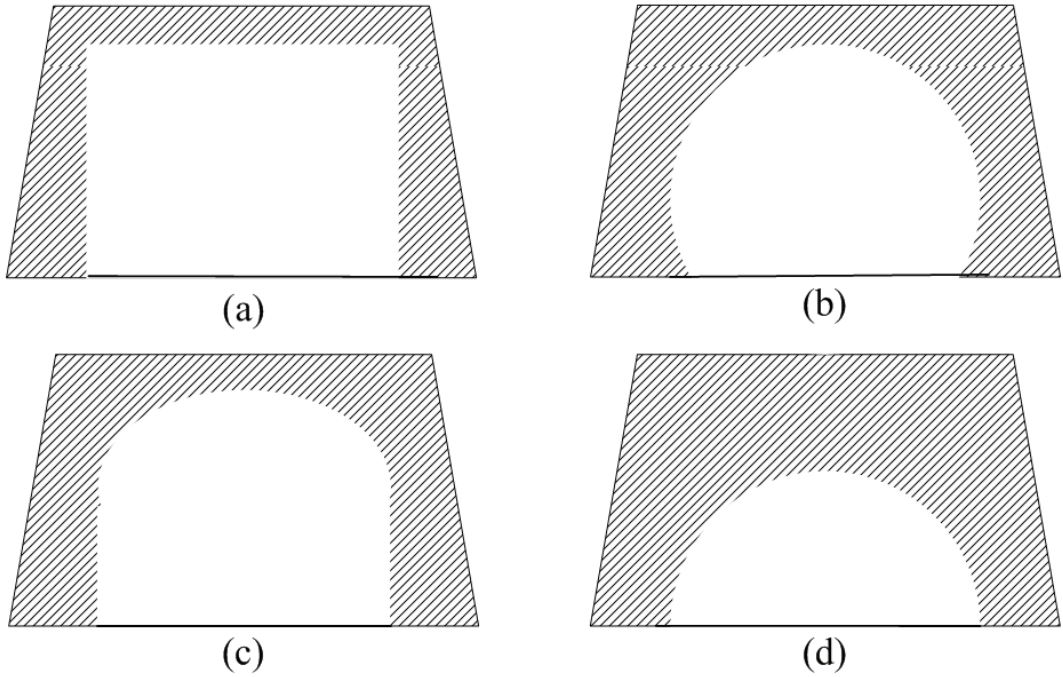}
\caption{Typical shapes of cross sections for tunnels.}
\label{fig1}
\end{figure}

The surface roughness and electromagnetic properties of tunnel walls also affect the radio propagation in tunnels. In \cite{Zho16_SR}, the influence of surface roughness has been studied. It is found that surface roughness of tunnel walls introduces additional power attenuation to radio signals. Moreover, the influence of humidity in conductivity and dielectric constant has been considered \cite{Che12_Hum}. The results show that there is negligible effect on the dielectric constant, but not on conductivity.
Beyond that, the position, polarization, and radiation pattern of transmitter and receiver antennas will also influence the radio wave propagation along the tunnel\cite{Sun10_mine}, \cite{Mar94_AP}. In\cite{Huo09_AP} and \cite{Han12_AP}, the authors provide the optimal radiation pattern and position of antenna inside the tunnel. By using the antennas with appropriate radiation pattern at an appropriate position, the attenuation of radio signals inside tunnel can be reduced. For investigating the influence of the antenna directivity on PL and time dispersion, the directional and omnidirectional antennas are considered in \cite{Ris12_AP} under the LoS and NLoS underground environments. The results show that the omnidirectional antennas can offer better signal coverage in NLoS tunnel environment, while the directional antennas can reduce the time dispersion parameters to acquire a better channel capacity \cite{Ris12_AP}. The polarization of the transmitter and receiver antennas has been studied in \cite{Ker00_AP}. In an empty straight rectangular tunnel environment, rms
delay spread of horizontally polarized transmitter and receiver antennas is greater than that of vertically polarized transmitter and receiver antennas if the tunnel width is larger that the tunnel height. In addition, the attenuation of EM wave for the horizontal polarization is lower than that for the vertical polarization.

\section{HST Channel Models in Tunnel Scenarios}\label{Sec3}
Several HST tunnel channel models have been presented in the literature \cite{Zhan04_DL, Zhe09_GBSB, Avazov13_c2c, Ber11_CM, Wan14_CM, Yao11_CM, Ye16_CM, Ran16_CM}. In this section, we will first discuss possible network architectures for tunnel scenario and then present various types of tunnel channel models according to different modeling methods.

\subsection{Network architectures for tunnels}
As mentioned earlier, two solutions are introduced to provide wireless coverage inside tunnels, i.e., leaky feeders and DAS \cite{Gua12_Leakey}. Leaky feeders are widely used in current HST tunnel communications, as they can provide a good coverage and do not require special planning. However, it requires a high cost of the installation and a rather complex maintenance, especially when medium-length or even long tunnels are needed in newly-built high speed railways. In this case, solutions based on the use of antennas are becoming more attractive, such as the DAS. In DAS, all the antenna elements are installed at planned distance intervals and connected to a BS via wires or fibers. Compared with leaky feeders, the DAS can provide considerable gain in coverage, capacity, and spatial diversity against fading by using antenna elements at different locations \cite{Bri07_DAS}. It also has some other advantages, such as quick installation and easy maintenance. Moreover, adopting the conventional cellular architecture, where the users inside trains communicate directly with outdoor base stations (BSs), leads to  several communication problems. Therefore, MRS need to be considered. It can be deployed on the surface of the train to improve the quality of received signals \cite{Gha15_HST}, and used to solve the spotty coverage problem and reduce the penetration loss of signals. In addition, potential applications for MIMO technique to increase the channel capacity of the propagation channel in tunnels are investigated \cite{Par09_MIMO}, \cite{Lie03_MIMO}. The appropriate combination of DAS, MRS, and MIMO technique, as illustrated in Fig.~2, is viable to meet the continuous and high-quality wireless communication requirements inside the tunnel. 
Furthermore, considering other cellular architectures in the future, more HST tunnel channel models are needed.
 
\begin{figure}[!t]
\centering
\includegraphics[width=5in]{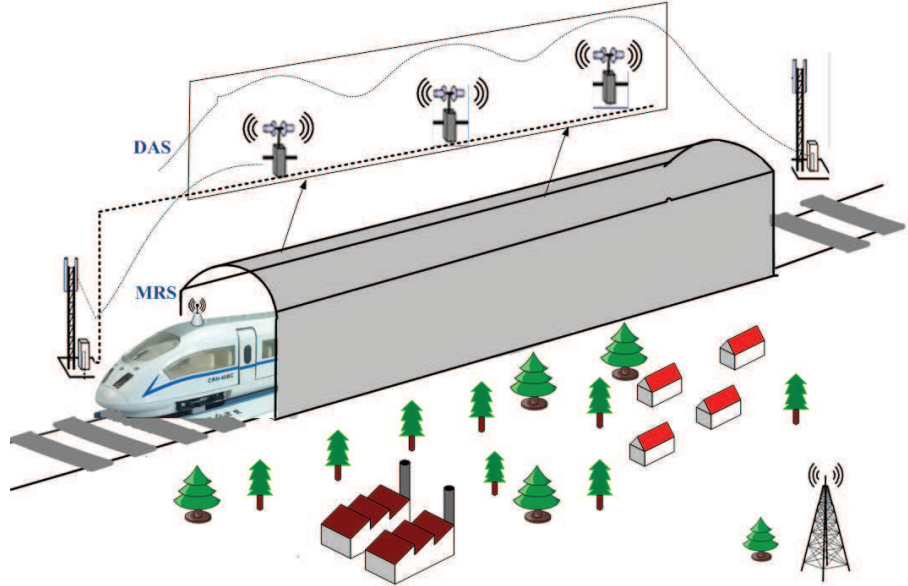}
\caption{HST cellular architecture for tunnel scenario.}
\label{fig2}
\end{figure}
\subsection{Modeling approaches of HST tunnel channel models}
According to different modeling approaches, the current tunnel channel models in the literature \cite{Min11_GO, Mas09_GO, Cho11_GO, Hai12_Sub, Liu15_MM, Gen12_CM, Che16_CM, Arg14_CM, Liu12_CM, Zha16_PG, Zho15_CM, Zha16_CM, Ge15_Mark, Mao09_Graph}, presented in Table~\ref{table_channel_models}, can be classified as deterministic \cite{Ems75_mine}, \cite{Taf05_FDTM} and stochastic channel models \cite{Zhe09_GBSB}, \cite{Avazov13_c2c}, \cite{Hai12_Sub}. The detailed classification of tunnel channel models is illustrated in Fig.~3.
\begin{figure}[!b]
\centering
\includegraphics[width=5in]{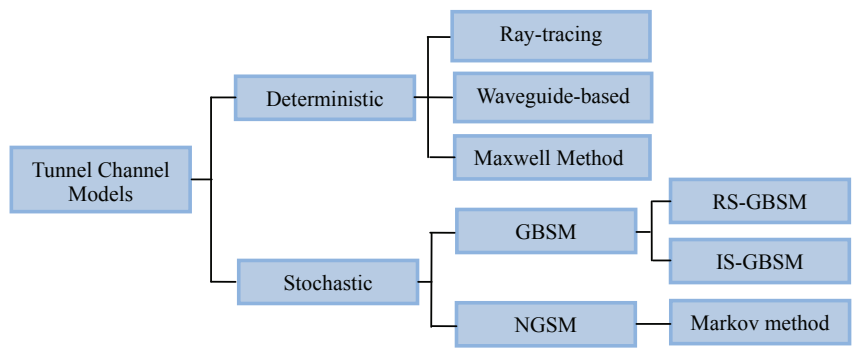}
\caption{Classification of HST tunnel channel models.}
\label{fig3a}
\end{figure}
%\cite{Ber11_CM}{Yao13_CM}{Wan13_CM}{Liu12_CM}{Hai12_CM}{Gen12_CM}

\begin{table}
	\center
	    \caption{Important tunnel channel models.}
    \begin{tabular}{| c | c | c | c | c |}
    \hline
\textbf{Ref.} &
\textbf{Channel Model} &
\textbf{Scenario} &
\textbf{Channel Characteristics}&
\textbf{Antenna Config.}\\
\hline
\cite{Gen12_CM} & Ray-tracing model &
$\begin {matrix} 
 \textrm{Rectangular} \\
 \textrm{tunnel} 
 \end {matrix} $
& the received power
& SISO \\
\hline
\cite{Che16_CM} & Ray-tracing model &
$\begin {matrix}
 \textrm{Rectangular} \\
 \textrm{subway tunnel}
 \end {matrix} $
&
$\begin {matrix} 
\textrm{PSD,} \\
\textrm{Doppler spread,}\\
\textrm{Doppler shift}
\end {matrix} $
& SISO \\
\hline
\cite{Sun10_mine} & Multi-mode model &
$\begin {matrix} 
\textrm{Rectangular}\\
\textrm{road tunnel,}\\ 
\textrm{subway tunnel}
\end {matrix} $
& $\begin {matrix} \textrm{field distribution,}\\
\textrm{PDP}
\end {matrix} $
& SISO \\
\hline
\cite{Arg14_CM} &
$\begin {matrix} \textrm{Multi-mode}\\
\textrm{waveguide model}
\end {matrix} $
&
$\begin {matrix} 
\textrm{Rectangular}\\
\textrm{underground mine,}\\
\textrm{Semicircular}\\
\textrm{subway tunnel}
\end {matrix} $
& $\begin {matrix} \textrm{angular properties,}\\
\textrm{correlation of array elements,}\\
\textrm{PAS}
\end {matrix} $
& MIMO \\
\hline
\cite{Liu12_CM} & GO model &
$\begin {matrix} 
\textrm{Rectangular}\\
\textrm{underground mine}
\end {matrix} $
&
$\begin {matrix} \textrm{large-scale fading,}\\
\textrm{small-scale fading}
\end {matrix} $
& SISO \\
\hline

\cite{Wan14_FSMM} & FSMM & 
$\begin {matrix} \textrm{Rectangular}\\
\textrm{subway tunnel}
\end {matrix} $
& $\begin {matrix} \textrm{number of states,}\\
\textrm{distance interval,}\\
\textrm{SNR}
\end {matrix} $
& SISO \\
\hline

\cite{Zha16_PG} & $\begin {matrix} \textrm{Propagation
-graph}\\
\textrm{theory based model}
\end {matrix} $
&
Arched tunnel
& $\begin {matrix} \textrm{channel coefficients,}\\
\textrm{CIR in delay,}\\
\textrm{antennas’ correlation coefficient,}\\
\textrm{channel capacity}
\end {matrix} $
& MIMO \\
\hline
\cite{Zho15_CM} & $\begin {matrix} \textrm{Physics-based}\\
\textrm{deterministic UWB}
\end {matrix} $
&
Rectangular Tunnel
& $\begin {matrix} \textrm{received power,}\\
\textrm{rms delay spread,}\\
\textrm{CIR,}\\
\textrm{channel transfer function}
\end {matrix} $
& SISO \\
\hline

\cite{Zhe09_GBSB} & GBSB model & 
Rectangular Tunnel &
$\begin {matrix} \textrm{space-time correlation function, } \\
\textrm{PDF of AoA, Rice factor
} \end {matrix} $
& MIMO \\
\hline

\cite{Hai12_Sub} & WINNER model & 
$\begin {matrix} \textrm{Rectangular} \\
\textrm{Subway tunnel}
\end {matrix} $
 &
$\begin {matrix} \textrm{PL, fast fading,} \\
\textrm{delays, AoA, AoD}
\end {matrix} $
& MIMO \\
\hline
%
%\cite{Ber11_CM} & GBSM & Tunnel &
%$\begin {matrix} \textrm{rms delay, } \\
%\textrm{Doppler spreads,} \\
%\textrm{excess delay, } \\
%\textrm{maximum Doppler dispersion,} \\
%\textrm{PDP, PSD}
%\end {matrix} $
%& MIMO \\
%\hline
\cite{Avazov13_c2c} & GBSM &
$\begin {matrix} \textrm{Semicircular} \\
\textrm{tunnel}
\end {matrix} $
&
$\begin {matrix} \textrm{time-variant transfer function,} \\
\textrm{frequency correlation function,}\\
\textrm{CCF, ACF}
\end {matrix} $
& MIMO \\
\hline
\cite{Zha16_CM} & Hybrid
model
 & Rectangular tunnel &
The received power
& SISO \\
\hline

\multicolumn{5}{| l |}{$\begin {matrix} \textrm{AoA: angle of arrival; AoD: angle of departure; rms: root mean square; PSD: power spectrum density;}\\
\textrm{ PAS: power azimuth spectrum}
\end {matrix} $
} \\
\hline
    \end{tabular}
	\label{table_channel_models}
\end{table}

\subsubsection{Ray-tracing channel model}
Ray-tracing technique has widely been used in predicting the radio wave propagation in confined environments like HST tunnels. Ray-tracing channel model can be applied to predict the PL and the signal delay inside tunnels at any location of the receiver \cite{Sun10_mine}. Based on the GO theory and uniform-theory-of diffraction (UTD), the EM waves are regarded as optical rays reflected from the tunnel walls and diffracted near the tunnel edges. However, as a special type of indoor scenario, the tunnel environment consisting of not only the walls, but also some other obstacles, such as rails, devices and the moving trains. Since the surfaces of these obstacles are not perfectly flat, ray diffusion should be considered, and the room EM wave propagation can be applied to analyze the diffuse field. Based on the room EM theory, models of the delay power spectrum of confined room channels have also been proposed. At the receiver, the EM field is obtained by the summation of the direct ray and diffused ones. The different phases of the summed rays will result in a variation of signal power along the distance, and the signal propagation predictions in tunnels can be developed. In \cite{Cic96_RT}, based on a new ray launching method, a 3D ray-tracing channel model in HST tunnel scenarios was introduced. This model resulted in a complex CIR that incorporates some channel information, such as the waveguide effects in tunnels and the impact of  another passing train simulations for time delay and Doppler shift.

The ray paths can be calculated by several approaches, such as the images method, shooting and bouncing ray (SBR) method, and the ray-density normalization (RDN) method. For the images method, all the reflected rays obtained at the receiver can be taken as radiated directly from the virtual sources, which can be obtained by the mirror symmetry of the transmitter. A ray-tracing model based on the images method was presented in \cite{Sun10_mine} to predict the rms delay spread in a tunnel. For the SBR method, the transmitter inside the tunnel is considered as a source that shoots a large number of rays in arbitrary directions. The received signal can be obtained by the summation of all contributions within a reception sphere, the radius of which depends on the length and angle of rays. The SBR-based ray-tracing method was proposed in \cite{Che96_SBR} to calculate the extra losses of tunnel curve, according to the number of reflections in the horizontal and vertical tunnel walls. For the RDN method, the radio waves that travel inside a tunnel have many propagation paths, while each propagation path is assumed to consist of several rays. The number of rays can be determined by the ray density and can be applied to normalize the contribution of each ray to the total field. The signals at the receiver are considered as the summation of all the rays with different amplitudes, phases, and ray densities. The RDN-based ray-tracing method can be used to calculate PL in arbitrary shaped tunnels \cite{Did_RDN}.

For the ray-tracing model based on GO theory, the EM fields at any point in space can be computed as a summation of rays from all possible paths. The paths are obtained using the method of images on the ceiling, floor and tunnel side walls. Thus, the electric field is computed by taking into account the laws of reflection and the constitutive parameters of the tunnel walls as follows \cite{Sun10_mine}:
\begin{equation}
E_x^{{R_x}} = E_x^{{T_x}}\sum\limits_{p,q} {\frac{{{e^{ - jk{r_{pq}}}}}}{{{r_{pq}}}}{S^p} \cdot {R^q}}
\end{equation}
where $E_x^{{T_x}}$ and $E_x^{{R_x}}$ are the electric fields at the transmitter and the reciter, respectively, ${r_{pq}}$ is the distance between the image and the receiver, $R^q$ and $S^p$ are the Fresnel reflection coefficients on the horizontal and vertical walls, respectively.
\subsubsection{Waveguide channel model}
Considering the geometry of tunnel and the conductivity of tunnel materials, the radio waves propagate inside a tunnel can be modeled as the same way as propagating inside a waveguide. As mentioned in \cite{Dud_WG}, when the frequency is higher than hundreds MHz, the waveguide effect will emerge. In addition, due to the unique structure of tunnel, there are rich reflections and scattering components which will introduce the waveguide effect inside tunnel \cite{Ai_Scenarios12}. In \cite{Ems75_mine}, a waveguide model was proposed, which adopts the modal theory to describe the radio wave propagation inside a tunnel that is considered as a rectangular waveguide. The mode, also called the transverse mode, is used to describe the field distribution of waveguide cross section. There are two kinds of propagation modes propagating in a waveguide: transverse electric (TE) mode (or $H_{mn}$) and transverse magnetic (TM) mode (or $E_{mn}$). Each mode has a cutoff frequency $f_{T}$, which is related to the tunnel size and mode values $m$ and $n$. When a given operating frequency $f_{c}$ is higher than $f_{T}$ of one or more modes, these modes can exist inside a tunnel. The field distribution can be viewed as the weighted sum of all modes field. In the near region, the EM field consists of many modes, which interact and result in the rapid attenuation. There are many factors contributing to the signal attenuation in tunnels, such as the operating frequency, tunnel size, permittivity of tunnel walls, and propagation modes. In the far region, the lowest-order mode is dominant. A waveguide model can be applied to model the far region in tunnels with good approximation, while it is not suitable to approximate the signal propagation in the near region. Therefore, a waveguide model should be combined with another model, which can model the multi-mode cases, to model a completed HST tunnel channel.

With the increase of the communication frequency, the operating frequency can be higher than the cutoff frequencies of many propagation modes. Thus, there will exist a wide range of modes propagating inside tunnels. In long tunnels, when the operating frequency is up to a few GHz, the distance of near region becomes longer, and therefore the time duration that train encounters in the near region becomes longer. This means that the near region will become larger with the increase of the operating frequency inside tunnels. Moreover, when a train travels inside a tunnel, the train itself will also have an influence on the field distribution. Hence, the multi-mode wave propagation and the impact of train itself on the field distribution at higher operating frequency need to be further investigated.
\subsubsection{Full-wave model}
The full-wave model, such as finite-difference time-domain (FDTD) technique \cite{Taf05_FDTM}, can be obtained by solving Maxwell equations using numerical methods. There are several numerical methods to be applied to solve the Maxwell equations. The most often used ones are FDTD, method of moments (MoM) \cite{Gib_MoM}, finite element method (FEM) \cite{{Poi_FEM}}, and vector parabolic equation (VPE) method \cite{Hro14_Sur}, \cite{Pop_VEP}. 他The FDTD technique focuses on solving partial differential equations at discrete times and discrete points. It can be applied to study the EM propagation accurately in complex environments as it fully considers the influences of reflection, diffraction, and refraction. The MoM is a widely used approach that can solve scattering, EM boundary, and volume integral equation problems. By using MoM, the operator equations can be expressed in a matrix form and EM field can be obtained by solving the matrix. The FEM is often used to find approximate solutions for partial differential equations. It can be applied to calculate EM distributions in arbitrary-shaped railway tunnels, but it has high computational complexity. This method can be used to analyze the electromagnetic field distribution in railway tunnel scenario with train \cite{{Poi_FEM}}. The VPE method can be used to calculate the EM field in straight and curved tunnels with reasonable computational complexity \cite{Hro14_Sur}.

\subsubsection{Hybrid model}
A variety of approaches has been applied to model the propagation channel in tunnel scenarios. Each method has its own advantages and disadvantages. To achieve the complementary advantages, hybrid channel modeling methods have been investigated. In \cite{Sun10_mine}, a multi-mode model was developed, which is a hybrid model that combines a GO model and a waveguide model using Poisson sum formulas. It used a mode matching technique to convert sum of rays of the GO model to sum of modes by mode intensities, and can characterize the natural wave propagation completely both in near and far regions of the source. Moreover, the PDP can also be characterized, which is related to the dispersion among modes and frequency elements. Further, based on the multi-mode model, an in-depth analysis of the tunnel channel characteristics was presented \cite{Sun10_mine}. %It is applicable to the railway tunnel scenarios to predict the PL as well as the delay spread.
In \cite{Liu15_MM}, a time-dependent multi-mode model was proposed and some small-scale fading characteristics were further investigated, such as the temporal ACF and Doppler PSD. Moreover, the received power along the multi-mode and single-mode cases are shown in Fig. 3. From this figure, we can observe that the lowest-order mode experiences the least attenuation and higher-order modes experience higher attenuation.
\begin{figure}[hbtp]
\centering
\includegraphics[width=5in]{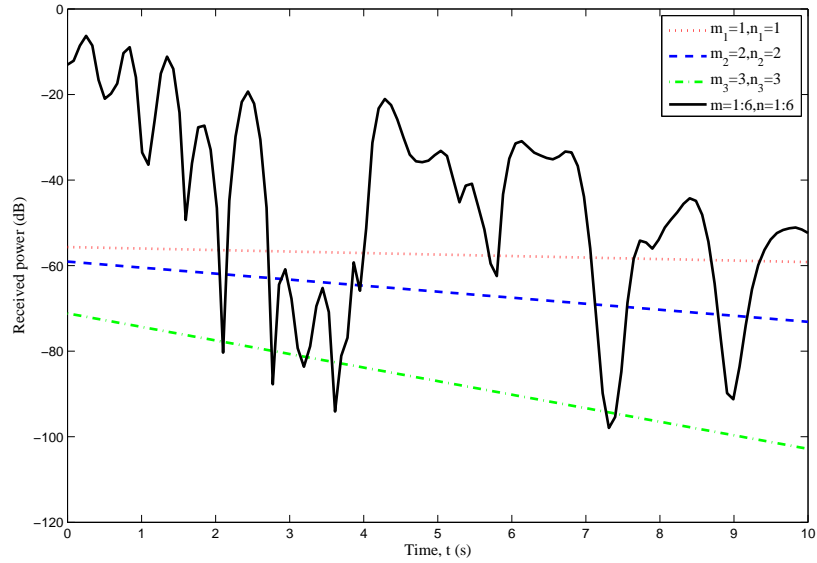}
\caption{The received power for multi-mode and single-mode cases in a multi-mode tunnel channel model.}
\label{Fig3}
\end{figure}
In addition, a hybrid model, which combines the ray-tracing and VPE method, is proposed in \cite{Zha16_CM}. By using the advantages of VEP, the limitation of ray-tracing method can be compensated. Therefore, the appropriate combination of two models, such as GO model with waveguide model, and ray-tracing with full-wave models, can be considered in the future modeling.

\subsubsection{GBSM}
A GBSM can be characterized by the specific transmitter, receiver, and scatterer geometries which are assumed to follow certain probability distributions \cite{Gha12_HST}. In RS-GBSMs, all the effective scatterers are assumed to be located on regular shapes, such as two-dimensional (2D) one-ring, two-ring, and ellipse models, and 3D one-sphere, two-sphere, and elliptic-cylinder models. Based on the relationships of geometrical shapes, the CIR can be derived and channel statistics can be further calculated \cite{Gha15_IMT}. In \cite{Gha15_HST}, a non-stationary wideband RS-GBSM for HST channels was proposed. The proposed GBSM is based on a concentric multi-ellipse model with all the model parameters as time-variant. Some small-scale fading characteristics were derived, such as the temporal ACF, spatial CCF, and Doppler PSD. It has been shown that the time-varying angles will affect the time-variant space CCFs, ACFs, and Doppler PSDs. Note that the Doppler PSD is symmetrical for isotropic cases only, and different angular parameters, such as angle of motion of HST and the initial mean AoA will affect considerately the trends of PSDs. The GBSM in \cite{Gha15_HST}, \cite{Gha15_GenerCM} can be applied to different HST scenarios, such as the open space, viaduct, and cutting scenarios, but not for tunnel scenarios. Because of the long and narrow space inside a tunnel, the complex structure of tunnel walls, and poor smoothness of interior walls, a tunnel can bring more scatterers. The scatterers generally concentrate on the top, bottom, and both sides of tunnel walls. Therefore, the geometric distribution of scatterers in tunnels is very different from those in other HST scenarios. In \cite{Zhe09_GBSB}, a 2D narrowband geometry-based single-bounced (GBSB) channel model was proposed, which assumed that the scatterers are well-distributed on both sides of the tunnel. The CIR was expressed by the signal waves summations of different amplitudes, phases, and delays at the receiver. The proposed GBSB model is relatively simple and cannot describe the real tunnel channel. Therefore, a 3D channel model considering both the azimuth and elevation angle are needed in tunnel scenarios. In \cite{Avazov13_c2c}, a 3D GBSM for road tunnel \cite{Mao13_Roadtraffic} was proposed, then some key statistical properties were studied. However, this tunnel GBSM is under the wide-sense stationary assumption which is unreasonable and ignores the non-stationarity resulting from the fast movement of the transmitter and/or the receiver \cite{Yuan15_V2V}, \cite{Yuan14_V2Viso}. From the above, a 3D non-stationary channel model in HST tunnel scenarios, is still desirable.
\subsubsection{FSMM}
In \cite{Wan14_FSMM}, an FSMM for tunnel channels in a communication-based train control system was proposed based on real channel measurements where the locations of the train were known. The proposed FSMM was characterized by channel states which can be defined according to different received SNR levels. Different from other existing tunnel channel models, the proposed FSMM takes the train locations into consideration, which makes the model more accurate. The tunnel can be divided into intervals in terms of the distance. Each interval is related to a state transition probability matrix and then an FSMM for tunnel channels can be designed. It has been demonstrated that the number of states has a certain influence on the accuracy of the proposed FSMM, as well the distance between the transmitter and receiver.

\section{Research Directions in HST Tunnel Channel Measurements and Models}\label{Sec4}
In this section, we will discuss a few future research directions in HST tunnel channel measurements and models, which can be helpful for carrying out future channel measurements and developing realistic tunnel channel models.
\subsection{Statistical properties}
For better understanding and analyzing of the HST communication system in tunnels, the studies of the statistical properties are essential. In Table~\ref{table_channel_measurement}, some channel characteristics were obtained from channel measurements. However, most of the characteristics are mainly focus on large-scale fading. In Table~\ref{table_channel_models}, some tunnel channel models have been proposed. However, the corresponding analysis of small-scale fading characteristics are simplified. They can not be applied to mimic the propagation environment inside tunnel very well. Hence, it is desirable to further study the statistical properties of HST tunnel channel models.
\subsection{Non-stationarity of HST tunnel channels}
Measurements have demonstrated that the stationary interval of HST channel can retain a very short time in \cite{Che12_SI, Wan09_SI, Mol09_SI}. The finding applied equally in the HST tunnel channels. However, few channel measurements and models in tunnel scenarios have considered the non-stationarity features. Therefore, non-stationary channel models considering the time-variant parameters should be further investigated, and ideally verified by real-field measurements.
\subsection{3D GBSMs}
The existing GBSB model \cite{Zhe09_GBSB} mainly considered the 2D influences of the tunnel side-walls under the assumption of wide-sense stationary condition. It was a simplified 2D channel model without considering the elevation angle. More accurate 3D non-stationary wideband tunnel channel models are needed, which should consider the elevation angels and the influence of train itself, tunnel ground, and tunnel roof. It can be used to mimic the real tunnel channel more accurately. Combining the WINNER model method and tunnels' unique propagation characteristics, the HST tunnel surroundings can be characterized as a 3D regular shape, such as cuboid or circular model \cite{Liu16_TunGBSM}.
Fig. 4 illustrates the proposed 3D tunnel GBSM, which consists of the LoS, single bounce, and multi-bounce diffuse components. This kind of tunnel channel model can be developed under the clusters-based framework. It assumes the clusters are randomly distributed on the tunnel internal surfaces. Then, according to the geometrical relations of AoAs and AoDs, the CIR can be derived. Furthermore, the statistical properties can be further investigated, such as the temporal ACF, spatial CCF, and PSD for the HST channels in tunnel scenarios.

\begin{figure}[hbtp]
\centering
\includegraphics[width=5in]{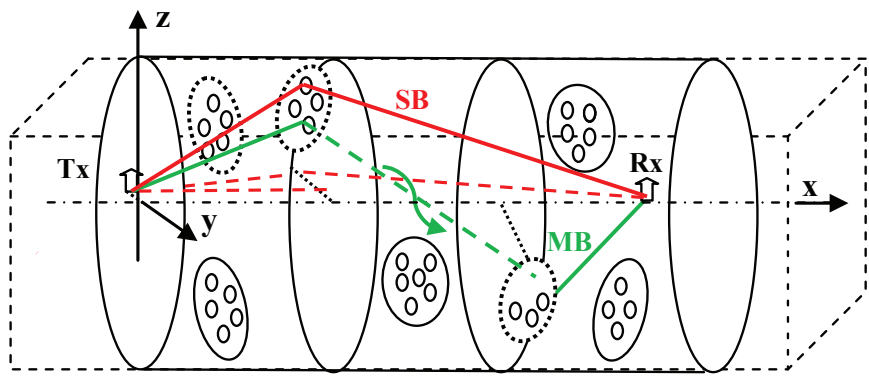}
\caption{A 3D RS-GBSM for HST tunnel scenarios.}
\label{Fig4}
\end{figure}
\subsection{Generic channel model for different types of tunnels}
There are several types of HST tunnels in reality, such as rectangular, circular, and arched tunnels. Different shapes of the tunnels have different impacts on channel characteristics, and will result in different degrees of attenuation of signals. In a rectangular tunnel, there are two vertical walls and two horizontal planes. In a circular tunnel, there are circular walls and a floor. Moreover, there are also two kinds of arched tunnels. One consists of an arched roof and three walls, and the other includes arched walls and a floor only. For different types of tunnels, there are different methods to model the underlying channel. However, an accurate generic channel model that can be applied to different types of tunnel channels by adjusting channel parameters is desirable and deserves further investigation in the future.
\subsection{System performance}
For system design and network planning, the investigation of HST tunnel communication system performance is essential. In \cite{Lie14_SP}, MIMO techniques have been investigated to improve link reliability using bit error rate (BER) and/or channel capacity. In addition, some alternative and easiest receiving diversity schemes, such as selection combining and maximum ratio combining (MRC), presented. 
In \cite{Mou10_SP}, the BER performance of a subway tunnel communication systems was
studied by considering the space diversity techniques, such as MRC and cyclic delay diversity. Furthermore, the implementation of DAS in railway tunnel communication systems was evaluated in \cite{Shu16_SP} by analyzing the coverage efficiency. In the future, some new technologies, such as Massive MIMO, can be applied to HSTs to boost their performance. Therefore, more performance analysis of HST tunnel communication systems, evaluating other schemes and considering more system performance indicators, e.g., capacity and
quality of service (QoS), is required. 

\section{Conclusions} \label{Sec5}
This article has provided a review of channel measurements and models in HST tunnel scenarios. We have surveyed tunnel channel measurements according to carrier frequencies, tunnel parameters, antenna configurations, and channel statistics. Then, we have classified some existing tunnel channel models according to different modeling methods. Some large-scale and small-scale fading characteristics have also been presented. Finally, to develop more practical HST channel models for 4G and 5G systems, some future research directions in HST tunnel channel measurements and modeling have been highlighted.

%%%%%%%%%%%%%%%%%%%%%%%%%%%%%%%%%%%%%%%%%%%%%%%%%%%%%%%
%%% Acknowledgements. 致谢
%%%%%%%%%%%%%%%%%%%%%%%%%%%%%%%%%%%%%%%%%%%%%%%%%%%%%%%
\Acknowledgements{The authors would like to acknowledge the support from the International S\&T Cooperation Program of China (No. 2014DFA11640), EU H2020 ITN 5G Wireless project (Grant No. 641985), EU FP7 QUICK project (Grant No. PIRSES-GA-2013-612652), EPSRC TOUCAN project (Grant No. EP/L020009/1), and the China Scholarship Council (Grant No. 201506450042).}
%%%%%%%%%%%%%%%%%%%%%%%%%%%%%%%%%%%%%%%%%%%%%%%%%%%%%%%
%%% Reference section. 参考文献
%%% citation in the content using "some words~\cite{1,2}".
%%% ~ is needed to make the reference number is on the same line with the word before it.
%%%%%%%%%%%%%%%%%%%%%%%%%%%%%%%%%%%%%%%%%%%%%%%%%%%%%%%

%%%%%%%%%%%%%%%%%%%%%%%%%%%%%%%%%%%%%%%%%%%%%%%%%%%%%%%
%%% Appendix sections. 附录章节, 非必选
%%%%%%%%%%%%%%%%%%%%%%%%%%%%%%%%%%%%%%%%%%%%%%%%%%%%%%%
%\begin{appendix}
%\section{Name}

%\end{appendix}


\begin{thebibliography}{1}
\bibitem{IMT_5G}
IMT-2020 Promotion Group. 5G visions and requirements. White Paper.
\bibitem{Wan14_5G}
Wang C X, Haider F, Gao X, et al. Cellular architecture and key technologies for 5G wireless communication networks. IEEE Commun. Mag., 2014,~52:~122--130.
\bibitem{Ai_Scenarios12}
Ai B, He R, Zhong Z D, et al. Radio wave propagation scene partitioning for high-speed rails. Int. J. Antennas Propag., 2012, ~2012:~1--7.
\bibitem{Gha15_HST}
Ghazal A, Wang C X, Ai B, et al. A non-stationary wideband MIMO channel model for high-mobility intelligent transportation systems. IEEE Trans. Intelligent Transportation Systems, 2015,~16:~885--897.
\bibitem{Fok10_Sur}
Fokum D T and Frost V S. A survey on methods for broadband internet access on trains. IEEE Commun. Surveys Tut., 2010,~12:~171--185.
\bibitem{Wang16_HST_Sur}
Wang C X, Ghazal A, Ai B, et al. Channel measurements and models for high-speed train communication systems: a survey. IEEE Commun. Surveys Tut., 2016,~18:~974--987.
\bibitem{Hro14_Sur}
Hrovat A, Kandus G, and Javornic T. A survey of radio propagation modeling for tunnels. IEEE Commun. Surveys Tuts., 2014,~16:~658--669.
\bibitem{Gua12_Leakey}
Guan K, Zhong Z D, Alonso J I, et al, Measurement of distributed antenna system at 2.4GHz in a realistic subway tunnel environment. IEEE Trans. Veh. Technol., 2012, 61:~834--837.
\bibitem{Bri07_DAS}
Briso-Rodriguez C, Cruz J M, and Alonso J I. Measurements and modeling of distributed antenna systems in railway tunnels. IEEE Trans. Veh. Technol., 2007,~56:~2870--2879.
\bibitem{Gua13_DAS}
Guan K, Zhong Z D, and Ai B. Statistic modeling for propagation in tunnels based on distributed antenna systems. Proceedings of AP-SURSI’13, Florida, USA, 2013.~1920--1921.
%\bibitem{Gua12_DAS}
%Guan K, Zhong Z D, Alonso J I, et al, Measurement of distributed antenna system at 2.4 GHz in a realistic subway tunnel environment. IEEE Trans. Veh. Technol., 2012,~61:~834--837.
\bibitem{Mah74_GO}
Mahmoud S F and Wait J R. Geometrical optical approach for electromagnetic wave propagation in rectangular mine tunnels. Radio Science., 1974,~9:~1147--1158.
\bibitem{Por02_GO}
Porrat D. Radio propagation in hallways and streets for UHF communications. Ph.D. thesis. California: Stanford University, 2002.
\bibitem{Zha16_GO}
Zhang J C, Tao C, Liu L, et al. A study on channel modeling in tunnel scenario based on propagation-graph theory. Proceedings of VTC'16-Spring, Nanjing, China, 2016.~1--5.
\bibitem{Wan98_GO}
Wang Y H, Zhang Y P, and Kouyoumjian R G. Ray-optical prediction of radio-wave propagation characteristics in tunnel environments part 1: theory, part 2: analysis and measurements. IEEE Trans. Antennas Propag., 1998,~46:~1328--1345.
\bibitem{For14_WG}
Forooshani A E, Noghanian S and Michelson D G. Characterization of angular spread in underground tunnels based on the multimode waveguide model. IEEE Trans. Commun., 2014,~62:~4126--4133.
\bibitem{Dud05_WG}
Dudley D G. Wireless propagation in circular tunnels. IEEE Trans. Antennas Propag., 2005,~53:~435--441.
\bibitem{Did01_WG}
Didascalou D, Maurer J, and Wiesbeck W. Subway tunnel guided electromagnetic wave propagation at mobile communications frequencies. IEEE Trans. Antennas Propag., 2001,~49:~1590--1596.
\bibitem{Zha97_WG}
Zhang Y P and Hwang Y. Enhancement of rectangular tunnel waveguide model. Proceedings of APMC'97, Hong Kong, China,~1997.~197--200.
\bibitem{Ems75_mine}
Emslie A G, Lagace R L, and Strong P F. Theory of the propagation of UHF radio waves in coal mine tunnels. IEEE Trans. Antennas Propag.,1975,~23:~192--205.
\bibitem{Ran12_Maxwell}
Rana M and Mohan A. Segmented-locally-one-dimensional-FDTD method for EM propagation inside large complex tunnel environments.  
IEEE Trans. Mag., 2012,~48:~223--226.
\bibitem{Taf05_FDTM}
Taflove A and Hagness S C. Computational electrodynamics: the finite-difference time-domain method, 3rd edition. Norwood. MA: Artech House, 2005.
\bibitem{Wan00_Maxwell}
Wang Y, Safavi-Naeini S, and Chaudhuri S. A hybrid technique based on combining ray tracing and FDTD methods for site-specific modeling of indoor radio wave propagation. IEEE Trans. Antennas Propag., 2000,~48:~743--754.
\bibitem{Wan14_FSMM}
Wang H W, Yu F R, Zhu L, et al. Finite-state markov modeling for wireless channels in tunnel communication-based train control systems. IEEE Trans. Intell. Transp. Syst.,  ~2014,~15:~1083--1090.
\bibitem{Aik98_Mea}
Aikio P, Gruber R, and Vainikainen P. Wideband radio channel measurements for train tunnels. Proceedings of VTC'98, Ottawa, Canada, 1998.~460--464.
\bibitem{Gua15_Cur}
Guan K, Ai B, Zhong Z D, et al. Measurements and analysis of large-scale fading characteristics in curved subway tunnels at 920 MHz, 2400 MHz, and 5705 MHz. IEEE Trans. Intell. Transp. Syst.
 2015,~16:~2393--2405.
\bibitem{Zha03_PL}
Zhang Y P. A novel model for propagation loss prediction in tunnels. IEEE Trans. Veh. Technol., 2003,~52:~1308--1314.
\bibitem{Kim02_PL}
Kim Y M, Jung M S, Chin Y O, et al. Analysis of radio-wave propagation characteristics in curved tunnel. Electromagnetic Engineering Society, 2002,~13:~1017--1024.
\bibitem{He11_mea}
He R S, Zhong Z D, and Briso C. Broadband channel long delay cluster measurements and analysis at 2.4GHz in subway tunnels. Proceedings of VTC'11-Spring, Yokohama, Japan,~2011.~1--5.
\bibitem{Par08_LS}
Pardo J M G, Lienard M, Nasr A, et al. Wideband analysis of large scale and small scale fading in tunnels. Proceedings of ITST'08, Phuket, Thailand,~2008.~270--273.
\bibitem{Lie03_MIMO}
Lienard M, Degauque P, Baudet J, et al. Investigation on MIMO channels in subway tunnels. IEEE J. Sel. Areas Commun., 2003,  ~21:~332--339.
%\bibitem{For14_AS}
%A. E. Forooshani, S. Noghanian and D. G. Michelson, ``Characterization of Angular Spread in Underground TunnelsBased on the Multimode Waveguide Model," {\it IEEE Trans. on Commun.}, vol.~62, no.~11, pp. 4126--4133, Nov. 2014.
\bibitem{Cai16_Mea}
Cai X, Yin X F, Cheng X, et al. An empirical random-cluster model for subway channels based on passive measurements in UMTS. IEEE Trans. Commun., 2016,~64:~3563--3575.
\bibitem{Jia15_Mea}
Jia Y L, Zhao M, Zhou W Y, et al.  Measurement and statistical analysis of 1.89 GHz radio propagation in a realistic mountain tunnel.  Proceedings of WCSP'15, Nanjing, China,~2015.~1--5.
\bibitem{Zha15_Mea}
Zhang L, Fernandez J, Briso-Rodriguez C, , et al. Broadband radio communications in subway stations and tunnels. Proceedings of EuCAP'15, Lisbon, Portugal,~2015.~1--5.
\bibitem{Li15_Mea}
Li J X, Zhao Y P, Zhang J, et al. Radio channel measurements and analysis at 2.4/5 GHz in subway tunnels. China Commun., 2015,~12:~36--45.
\bibitem{Zha00_PL}
Zhang Y P, Jiang Z R, Ng T S, et al. Measurements of the propagation of UHF radio waves on an underground railway train. IEEE Trans. Veh. Technol., 2000,~49:~1342--1347.
\bibitem{Par08_RP}
Molina-Garcia-Pardo J M, Lienard M, Nasr A, et al. On the possibility of interpreting field variations and polarization in arched tunnels using a model for propagation in rectangular or circular tunnels. IEEE Trans. Antennas and Propag., 2008,~56:~1206--1211.
\bibitem{Kim03_PL}
Kim Y M, Jung M, and Lee B. Analysis of radio wave propagation characteristics in rectangular road tunnel at 800 MHz and 2.4 GHz. Proceedings of IEEE Antennas and Propag. Soc. Int. Symp., Columbus, USA,~2003.~1016--1019.
\bibitem{Bas14_Mea}
Bashir S. Effect of antenna position and polarization on UWB propagation channel in underground mines and tunnels. IEEE Trans. Antennas Propag., 2014,~62:~4771--4779.
\bibitem{Sav15_Mea}
Savic V, Ferrer-Coll J, Angskog P, et al. Measurement analysis and channel modeling for TOA-Based ranging in tunnels. IEEE Trans. Wireless Commun. 2015,~14:~456--467.
\bibitem{Li16_PL}
Li G K, Ai B, Guan K, et al. Path loss modeling and fading analysis for channels with various antenna setups in tunnels at 30 GHz band. Proceedings of EuCAP'16, Davos, Switzerland,~2016.~1--5.
\bibitem{He_TR}
He R S, Zhong Z D, Ai B, et al. Analysis of the relation between Fresnel zone and path loss exponent based on two-ray model. IEEE Antennas Wireless Propag. Lett., 2012,~11:~208--211.
\bibitem{Hro10_4S}
Hrovat A, Kandus G, and Javornik T. Four-slope channel model for path loss prediction in tunnels at 400 MHz. IET MicroW. Antennas Propag.,~2010,~4:~571--582.
%\bibitem{Zha97_BP}
%Y. P. Zhang and Y. Hwang, ``Enhancement of rectangular tunnel waveguide model," in {\it Proc. APMC}, HongKong, 1997, pp.~197-200.
\bibitem{Gua13_BP}
Guan K, Zhong Z D, Ai B, et al. Research of propagation characteristics of break point; near zone and far zone under operational subway condition. Wireless Personal Commun., 2013,~68:~489--505.
\bibitem{Mar51_Cutoff}
Marcuvitz N. Waveguide Handbook, New York, Toronto, London: Maraw-Hill. 1951.
\bibitem{Kwo04_RT}
Kwon H, Kim Y and Lee B. Characteristics of radio propagation channels in tunnel environments: a statistical analysis.  Proceedings of APS/URSI'04, Sendai, Japan, 2004.~2995--2998.
%\bibitem{Kwo04_PL}
%Kwon H, Kim Y M, and Lee B S. 
%Characteristics of radio propagation channels in tunnel environments: a statistical analysis. Proceedings of Antennas Propag. Symp., Sendai, Japan, ~2004.~2995--2998.
\bibitem{Mol09_TS}
Molina-Garcia-Pardo J M, Lienard M, and Degauque P. Propagation in tunnels: experimental investigations and channel modeling in a wide frequency band for MIMO applications. EURASIP J. Wireless Communications and Networking, 2009,~2009:~560--571.
\bibitem{Did01_TS}
Didascalou D, Maurer J, and Wiesbeck W. Subway tunnel guided electromagnetic wave propagation at mobile communications frequencies. IEEE Trans. Antennas Propag., 2001,~49:~1590--1596.
\bibitem{Che10_TS}
Cheng L and Zhang P. Influence of dimension change on radio wave propagation in rectangular tunnels. Proceedings of Wireless Communications, Networking and Mobile Computing, Beijing, China, ~2009.~1--3.
\bibitem{Wan10_TS}
Wang S. Radio wave attenuation character in the confined environments of rectangular mine tunnel. Modern Applied Science, 2010,~4:~65--70.
\bibitem{Cha10_TS}
Chang-sen Z and Li-fang G. Research on propagation characteristics of electromagnetic wave in tunnels with arbitrary cross sections. Proceedings of ICFCC'2010, Wuhan, China, 2010.~22--25.
\bibitem{Zho16_SR}
Zhou C M, and Jacksha R. Modeling and measurement of wireless channels for underground mines. Proceedings of APSURSI, Fajardo, Puerto Rico,~2016.~1253--1254.
\bibitem{Che12_Hum}
Cheng L, Zhang L, and Li J. Influence of mine tunnel wall humidity on electromagnetic waves propagation. Int. J. Antennas and Propag., 2012, ~2012:~1--5.
\bibitem{Sun10_mine}
Sun Z and Akyildiz I F. Channel modeling and analysis for wireless networks in underground mines and road tunnels. IEEE Trans. Commun., 2010,~58:~1758--1768.
\bibitem{Mar94_AP}
Mariage P, Lienard M, and Degauque P.  Theoretical and experimental
approach of the propagation of high frequency waves in road tunnels. IEEE Trans. Antennas Propag., 1994,~42:~75--81.
\bibitem{Huo09_AP}
Huo Y, Xu Z, Zheng H D, et al. Effect of antenna on propagation characteristics of electromagnetic waves in tunnel environments. Proceedings of PrimeAsia'09, Shanghai, China,~2009.~268--271.
\bibitem{Han12_AP}
Han X, Wang S, Fang T, et al. Propagation character of electromagnetic wave of the different transmitter position in mine tunnel. Proceedings of NSWCTC'09, Wuhan, China,~2009.~530--533.
\bibitem{Ris12_AP}
Rissafi Y, Talbi L, and Ghaddar M.  Experimental characterization of an UWB propagation channel in underground mines.  IEEE Trans. Antennas Propag., 2012,~60:~240--246.
\bibitem{Ker00_AP}
Kermani M and Kamarei M. A ray-tracing method for predicting delay spread in tunnel environments. Proceedings of  ICPWC'00, Hyderabad, India,~2000.~538--542.
\bibitem{Zhan04_DL}
Zhang Y P and Hong H J. Ray-optical modeling of simulcast radio propagation channels in tunnels. IEEE Trans. Veh. Technol., 2004,~53:~1800--1808.
\bibitem{Zhe09_GBSB}
Zheng H D and Nie X Y. GBSB model for MIMO channel and its spacetime correlation analysis in tunnel. Proceedings of NSWCTC'09, Wuhan, China,~2009.~1--8.
\bibitem{Avazov13_c2c}
Avazov N and Patzold M. A novel wideband MIMO car-to-car channel model based on a geometrical semi-circular tunnel scattering model. IEEE Trans. Veh. Technol., 2016,~65:~1070--1082.
\bibitem{Ber11_CM}
Bernado L, Roma A, Paier A, et al. In-tunnel vehicular radio channel characterization. Proceedings of VTC'11-Spring, Budapest, Hungary, ~2011.~15--18.
\bibitem{Wan14_CM}
Wang H W, Yu F R, Zhu L, et al. Finite-state markov modeling of tunnel channels in comminication-based train control systems. Proceedings of ICC’13, Budapest, Hungary, ~2013.~5047--5051.
\bibitem{Yao11_CM}
Yao S H, Wu X L. Modeling for MIMO wireless channels in mine tunnels.  Proceedings of IEEE ICEICE’11, Wuhan, China, 2011.~520--523.
\bibitem{Ye16_CM}
Ye X K, Cai X S, Wang H W, et al. Tunnel and non-tunnel channel characterization for high-speed-train scenarios in LTE-A networks. Proceeding of IEEE VTC’16-Spring, Nanjing, China,~2016.~1--5.
\bibitem{Ran16_CM}
Ranjany A, Misraz P, Dwivediz B, et al. Channel modeling of wireless communication in underground coal mines. Proceedings of  IEEE COMSNETS'16, Bangalore, India,~2016.
~1--2.
\bibitem{Par09_MIMO}
Molina-Garcia-Pardo J M, Lienard M, Stefanut P, et al. Modeling and understanding MIMO propagation in tunnels. J. Commun., ~2009,~4:~241--247.
\bibitem{Min11_GO}
Minghua J. A modified method for predicting the radio propagation characteristics in tunnels. Proceedings of WiCOM'11, Wuhan, China,~2011.~1--4.
\bibitem{Mas09_GO}
Masson E, Combeau P, Berbineau M, et al. Radio wave propagation in arched cross section tunnels simulations and measurements. J. Commun. 2009,~4:~276--283.
\bibitem{Cho11_GO}
Choudhury B and Jha R. A refined ray tracing approach for wireless communications inside underground mines and metrorail tunnels. Proceeding of IEEE AEMC'11, Kolkata, India,~2011.~1--4.
\bibitem{Hai12_Sub}
Hairoud S, Combeau P, and Pousset Y. WINNER model for subway tunnel at 5.8 GHz. Proceedings of IEEE ITST'12, Taibei, Taiwan, 2012.~743--747.
\bibitem{Liu15_MM}
Liu Y, Wang C X, Ghazal A, et al. A multi-mode waveguide tunnel channel model for high-speed train wireless communication systems. Proceedings of IEEE EuCAP'15, Lisbon, Portugal,~2015.~1--5.
\bibitem{Gen12_CM}
Gentile C, Valoit F, and Moayeri N. A retracing model for wireless propagation in tunnels with varying cross section. Proceedings of IEEE GLOBECOM'12, Anaheim, CA,~2012.~5027--5032.
\bibitem{Che16_CM}
Chen X, Pan Y T, Wu Y M, et al. Research on doppler spread of multipath channel in subwaytunnel. Proceedings of IEEE ICCP'14, Beijing, China,~2014.~56--59.
\bibitem{Arg14_CM}
Forooshani A E, Noghanian S, and Michelson D G. Characterization of angular spread in underground tunnels based on the multimode waveguide model. IEEE Trans. Commun., 2014,~62:~4126--4133.
\bibitem{Liu12_CM}
Liu C G, Chen Q, and Yang G W. A calculation model and characteristics analysis of radio wave propagation in rectangular shed tunnel. Proceedings of IEEE ISAPE'12, Xian, China, 2012.~535--539.
\bibitem{Zha16_PG}
Zhang J C, Tao C, Liu L, et al. A study on channel modeling in tunnel scenario based on propagation-graph theory. Proceedings of IEEE VTC'16-Spring, Nanjing, China,~2016.~1--5.
\bibitem{Zho15_CM}
Zhou C M. Physics-based ultra-wideband channel modeling fortunnel/mining environments. Proceedings of IEEE RWS'15, San Diego, USA, 2015.~92--94.
\bibitem{Zha16_CM}
Zhang X, Sood N, Siu J K, et al. A hybrid ray-tracing/vector parabolic equation method for propagation modeling in train communication channels. IEEE Trans. Antennas Propag., 2016,~64:~1840--1849.
\bibitem{Ge15_Mark}
Ge X, Tu S, Han T, et al. Energy efficiency of small cell backhaul networks based on Gauss-Markov mobile models. IET Networks,~2015,~4: ~158--167.
\bibitem{Mao09_Graph}
Mao G, Anderson B D O. Graph theoretic models and tools for the analysis of dynamic wireless multihop networks. Proceedings of WCNC’09, Budapest, Hungary,~2009.~1--6.
\bibitem{Cic96_RT}
Cichon D J, Zwick T, and Wiesbeck W. Ray optical modeling of wireless communications in high-speed railway tunnels. Proceedings of IEEE VTC'96-Spring, Atlanta, USA, 1996.~546--550.
\bibitem{Che96_SBR}
Chen S H and Jeng S. K. SBR image approach for radio wave propagation in tunnels with and without traffic. IEEE Trans. Veh. Technol., 1996,~45:~570--578.
\bibitem{Did_RDN}
Didascalou D, Schafer T, Weinmann F, et al. Ray-density normalization for ray-optical wave propagation modeling in arbitrarily shaped tunnels. IEEE Trans. Antennas Propag., 2000,~48:~1316--1325.
\bibitem{Dud_WG}
Dudley D, Mahmoud S, Lienard M, and Degauque P. On wireless communication in tunnels. Proceedings of IEEE APS/URSI'07, 2007.~3305--3308.
\bibitem{Gib_MoM}
Gibson W C. The method of moments in electromagnetics. Chapman \& Hall/CRC, Taylor \& Francis Group, 2008.
\bibitem{Poi_FEM}
Poitau G and Kouki A. Analysis of MIMO capacity in waveguide environments using practical antenna structures for selective mode excitation. Proceedings of CCGEI'04, Niagara, North America,~2004.~349--352.
\bibitem{Pop_VEP}
Popov A V and Zhu N Y. Modeling radio wave propagation in tunnels with a vectorial parabolic equation. IEEE Trans. Antennas Propag., 2000,~48:~1403--1412.
\bibitem{Gha12_HST}
Ghazal A,  Wang C X, Haas H, et al. A non-stationary geometry-based stochastic model for MIMO high-speed train channels. Proceedings of ITST’12, Taiwan,~2012.~7--11.
\bibitem{Gha15_IMT}
Ghazal A, Yuan Y,  Wang C X, et al. A non-stationary IMT-A MIMO channel model for high-mobility wireless communication systems. IEEE Trans. Wireless Commun. Accepted for publication.
\bibitem{Gha15_GenerCM}
Ghazal A, Wang C X, Liu Y, et al. A generic non-stationary MIMO channel model for different high-speed train scenarios. Proceedings of ICCC’09, Shenzhen, China,~2015.~1--6.
\bibitem{Mao13_Roadtraffic}
Mao R, Mao G. Road traffic density estimation in vehicular networks. Proceedings of WCNC’13, Shanghai, China,~2013.~4653--4658.
\bibitem{Yuan15_V2V}
Yuan Y, Wang C X, He Y, et al. 3D wideband non-stationary geometry-based stochastic models for non-isotropic MIMO vehicle-to-vehicle channels. IEEE Trans. Wireless Commun..~2015,~14. ~6883--6895.
\bibitem{Yuan14_V2Viso}
Yuan Y, Wang C X, Cheng X, et al. Novel 3D geometry-based stochastic models for non-isotropic MIMO vehicle-to-vehicle channels. IEEE Trans. Wireless Commun. 2014,~13.~298--309. 
\bibitem{Che12_SI}
Chen B, Zhong Z, and Ai B. Stationarity intervals of time-variant channel in high speed railway scenario. J. China Commun., 2012,~9:~64--70.
\bibitem{Wan09_SI}
Wang C X, Cheng X, and Laurenson D I. Vehicle-to-vehicle channel modeling and measurements: recent advances and future challenges. IEEE Commun. Mag., 2009,~47:~96--103.
\bibitem{Mol09_SI}
Molisch A F, Tufvesson F, Karedal J, et al. A survey on vehicle-to-vehicle propagation channels. IEEE Wireless Commun. Mag., 2009,~16:~12--22.
\bibitem{Liu16_TunGBSM}
Liu Y, Wang C X, Lopez C, et al. 3D non-stationary wideband circular tunnel channel models for high-speed train wireless communication systems. Sci China Inf Sci, 2016, accepted for publication.
\bibitem{Lie14_SP}
Lienard M, molina-Garcia-Pardo J M, Laly P, et al. Communication in tunnel: channel characteristics and performance of diversity schemes. Proceedings of URSI GASS'14, Beijing, China,~2014.~1--4.
\bibitem{Mou10_SP}
Mouaki B A, and Quenneville M. Performance evaluation of an L-band broadcast DAB/DMB system in simulated subway tunnel environment. Proceedings of VTC-Fall'10, Ottawa, Canada,~2010.~1--6.
\bibitem{Shu16_SP}
Shuo T L, Zhao K, and Wu H. Wireless communication for heavy haul railway tunnels based on distributed antenna systems. Proceedings of VTC'16-Spring, Nanjing, China,~2016.~1--5.

\end{thebibliography}
\end{document}